\documentclass[aps,prb,twocolumn,superscriptaddress,showpacs]{revtex4-1}

\usepackage{hyperref}
\usepackage[dvips]{graphicx}
\usepackage{amssymb,amsmath,amsfonts}
\usepackage[T1]{fontenc}
\usepackage{float}
\usepackage{color,soul}
\usepackage{booktabs}

\newcommand{\ke}{($\vec{k}, E$) }

\newcommand{\pin}{$P_{in}$}
\newcommand{\pout}{$P_{out}$}

\begin{document}


\title{Fe/GeTe(111) heterostructures as an avenue towards 'ferroelectric Rashba semiconductors' - based spintronics}


\author{Jagoda S\l awi\'{n}ska}
\affiliation{Consiglio Nazionale delle Ricerche, Istituto SPIN, UOS L'Aquila, Sede di lavoro CNR-SPIN c/o Univ. "G. D'Annunzio", 66100 Chieti, Italy}
\affiliation{Department of Physics, University of North Texas, Denton, TX 76203, USA}
\author{Domenico Di Sante}
\affiliation{Institut f\"{u}r Theoretische Physik und Astrophysik, Universit\"{a}t W\"{u}rzburg, Am Hubland Campus S\"{u}d, W\"{u}rzburg 97074, Germany}
\author{Sara Varotto}
\affiliation{Department of Physics, Politecnico di Milano, 20133 Milano, Italy}
\author{Christian Rinaldi}
\affiliation{Department of Physics, Politecnico di Milano, 20133 Milano, Italy}
\author{Riccardo Bertacco}
\affiliation{Department of Physics, Politecnico di Milano, 20133 Milano, Italy}
\author{Silvia Picozzi}
\affiliation{Consiglio Nazionale delle Ricerche, Istituto SPIN, UOS L'Aquila, Sede di lavoro CNR-SPIN c/o Univ. "G. D'Annunzio", 66100 Chieti, Italy}

\vspace{.15in}

\date{\today}

\begin{abstract}
By performing density functional theory (DFT) and Green's functions
calculations, complemented by X-ray Photoemission Spectroscopy, we
investigate the electronic structure of Fe/GeTe(111), a prototypical
ferromagnetic/Rashba-ferroelectric interface. We reveal that such system
exhibits several intriguing properties resulting from the complex
interplay of exchange interaction, electric polarization and spin-orbit
coupling. Despite a rather strong interfacial hybridization between Fe and
GeTe bands, resulting in a complete suppression of the surface states of
the latter, the bulk Rashba bands are hardly altered by the
ferromagnetic overlayer. This could have a deep impact on spin dependent
phenomena observed at this interface, such as spin-to-charge
interconversion, which are likely to involve bulk rather than surface
Rashba states.

\end{abstract}

\maketitle

\section{Introduction}
Ferroelectric Rashba semiconductors (FERSC) are a novel class of
relativistic materials whose bulk spin texture is intimately linked to
the direction of the ferroelectric polarization, thus allowing direct
electrical control over the spin degrees of freedom in a non-volatile
way.\cite{fersc, silvia, liebmann, minar1, minar2, fe-switching, calarco} Such property holds a large potential for
spintronics, or more specifically for spin-orbitronics,\cite{spin-orbitronics} aiming at injection, control
and detection of spin currents in non-magnetic materials.
While the Rashba effect has been mostly studied at surfaces
where inversion symmetry is intrinsically broken, in FERSC the so-called Rashba bulk bands (RB)
originate from inversion symmetry breaking due to the presence of a polar axis existing by definition in ferroelectrics.
Moreover, it has been predicted that the spin texture switches by changing the sign of polarization,
thus it can be reversed by electric field.

This fundamental prediction of spin texture switchability via
changing the sign of electric polarization ($\vec{P}$) has been recently
confirmed experimentally in the prototype material GeTe,\cite{gete_nano} representing a
first milestone towards the exploitation of the GeTe in spintronic devices, such as for
example the Datta-Das spin transistor.\cite{das, spintronics} However, the
design process of future applications requires a more detailed
characterization; due to the need of spin injection in any spintronics
devices, theoretical and experimental studies of GeTe-based interfaces
containing ferromagnets are particularly important. For this purpose, Fe
thin films seem to be a natural target material.\cite{bluegel}
Importantly, Fe/GeTe heterostructures have been recently realized
experimentally and have been shown to yield a spin-to-charge conversion
(SCC) in spin pumping experiments,\cite{fert} thus opening a realistic
perspective for the FERSC-based spintronics and making a need of further
theoretical input even more urgent.

In this paper, we employ density functional theory (DFT) to investigate
realistic Fe/GeTe interfaces, modeled by Te-terminated
$\alpha$-GeTe(111) surfaces capped by multilayer films of bcc Fe. As
mentioned above, the Fe layers on GeTe surfaces are interesting
for spin injection, but they can be also considered as a two-phase
multiferroic.\cite{schmid, fiebig, spaldin_science} Such composites have
been the subject of intensive studies in the last years given the
perspective of controling ferroelectricity (ferromagnetism) by magnetic
(electric) field due to the coupling between the magnetic and
ferroelectric properties in these materials. While Fe is a standard
ferromagnetic component considered in two-phase multiferroics,
Fe/BaTiO$_{3}$ being the prototype material,\cite{tsymbal_mae, tsymbal_prl,
tsymbal_strain, mertig, jap, radaelli, carmine} GeTe has never been considered as
a ferroelectric counterpart. Therefore, in order to clearly understand
the coupling mechanisms occuring at the interface, we will first analyze
the structural, electronic and magnetic properties of the interfacial
atoms, assuming different thicknesses of Fe films ranging from 1
monolayer (ML) to 6MLs. Such strategy, apart from providing essential
information about the magnetoelectric coupling, will also allow us to
identify when the interface properties become robust, an aspect
relevant for the design of novel GeTe-based devices. As a next step, we will
focus on the Fe/GeTe spin structure. The peculiar spin texture of bulk
and surface GeTe bands was studied in detail in our previous
works;\cite{fersc, liebmann, gete_nano} here, we will focus on the
influence of Fe on GeTe bulk Rashba bands and their hybridization. We
will analyze not only the dependence of the spin texture on the
thickness of the ferromagnetic film, but also on the electric polarization $\vec{P}$ which can
be parallel or anti-parallel to the surface's normal, and, finally, on
the magnetic anisotropy. Our theoretical analysis is complemented by
X-ray Photoemission Spectroscopy (XPS) measurements on Fe overlayer
deposited on (111)-oriented GeTe thin films.

\section{Methods}
\begin{figure}[h!]
\includegraphics[width=\columnwidth]{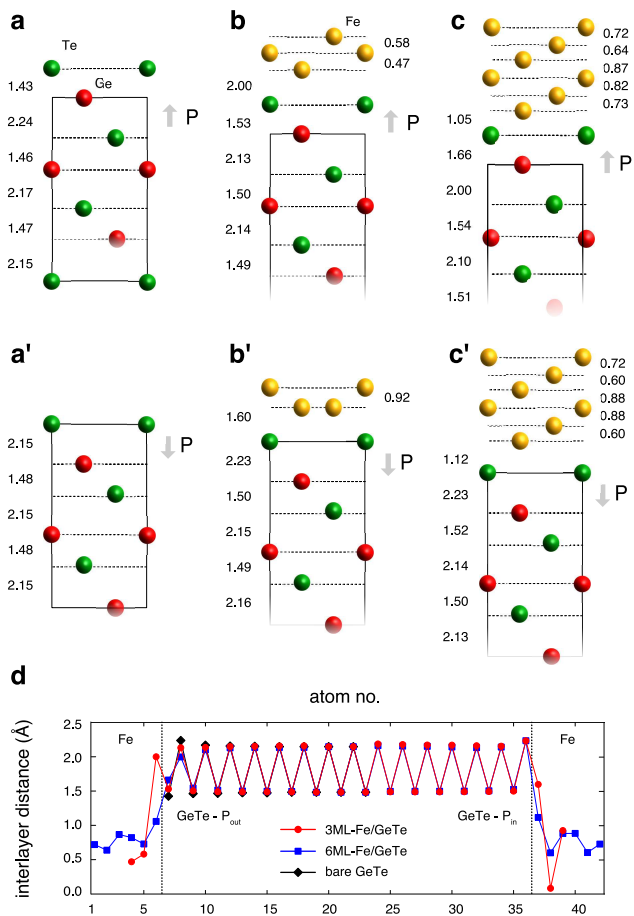}
\caption{\label{struct}
Schematic side view of optimized (a) $\alpha$-GeTe(111) (b) 3ML-Fe/GeTe(111) and (c) 6ML-Fe/GeTe(111) \pout\ surfaces. (a'-c') Same as (a-c) for \pin\ surfaces; in (a') the geometry is unrelaxed because the \pin\ surface turns out to be unstable. Te, Ge and Fe atoms are represented by green, red and yellow balls, respectively. Only the topmost surface layers are shown in each case. The primitive hexagonal bulk unit cells (marked by black rectangles) contain six atoms; in the surface calculations we use at least five such bulk blocks stacked along $z$ direction. Grey arrows denote the direction of $\vec{P}$. The interlayer distances are given in \AA. (d) Same interlayer distances plotted vs number of atomic layer. Our slabs by construction contain both \pout\ and \pin\ surfaces, therefore the left-hand (right-hand) side of the plot represent the interlayer distances of the former (latter), while the central part corresponds to constant values in the bulk GeTe. The interlayer distances in GeTe(111), 3ML-Fe/GeTe(111) and 6ML-Fe/GeTe(111) are plotted in black (diamonds), red (circles) and blue (square), respectively; note that, due to the fact that the relaxations never lead to \pin\ state within a bare GeTe(111) surface, the corresponding line ends in the bulk region. The \pin\ surface is omitted and only \pout\ surface is included.
}
\end{figure}

Te-terminated $\alpha$-GeTe(111) surface has been modeled using a
hexagonal supercell consisting of a sequence of 5 ferroelectric bulk
GeTe unit cells stacked along the $z$ axis.\cite{dronskowski} The slabs
contain one additional Te layer at the top surface which allowed us to
simultaneously study two different configurations, with dipole pointing
outwards (\pout) and inwards (\pin), represented by bottom and top side
of the slab, respectively (see Fig.\ref{struct} a-a'). As demonstrated
in our previous works,\cite{liebmann, gete_nano} for bare GeTe surfaces
only the \pout\ surface is stable, which can be rationalized recalling
that ferroelectric GeTe consists of an alternating long and short
Ge-Te bonds and the preferred termination corresponds to the breaking of
(weaker) long bonds; as a consequence the Te-terminated surface always
relaxes to the \pout\ configuration. Below, we present a detailed
characterization of the two configurations, as our results indicate that
the capping with Fe layers can stabilize both \pout\ and \pin\ phases.

The Fe/GeTe interfaces have been modeled assuming the pseudomorphic
matching between GeTe and bcc Fe(111) surfaces; this seems a reasonable
strategy given a relatively small mismatch of 4$\%$ between the in-plane
lattice parameter of GeTe surface (4.22 \AA) and the lattice constant of
bcc Fe (2.86 \AA). Moreover, recent LEED results clearly indicate the
hexagonal symmetry of the interface which further supports suitability
of our model.\cite{fert} Next, we consider different stacking orders of
Fe layers with respect to the substrate. The GeTe(111) hexagonal cell
contains three different high-symmetry sites to place the Fe atom: above
the topmost Te atom ($top$), above the topmost Ge atom ($hcp$) or above
the second Te atom ($fcc$). Stacking of two or more Fe layers can
arrange in six different configurations. We have considered all possible
stacking orders of 1ML, 2ML, 3ML and 6ML-Fe and further analyzed
properties of the most stable ones.

Our spin polarized DFT calculations
were performed using the Vienna Ab Initio Simulation Package (VASP)\cite{vasp1, vasp2}
equipped with the projector augmented-wave (PAW) method for electron-ion
interactions.\cite{paw1, paw2} The exchange-correlation interaction was
treated in the generalized gradient approximation in the parametrization
of Perdew, Burke, and Ernzerhof (PBE).\cite{pbe} In all simulations the
electronic wave functions were expanded in a plane-wave basis set of 400
eV, while the total energy self-consistency criterion was set to at
least $10^{-7}$ eV. The integrations over the Brillouin zone were
performed with ($10\times10\times1$) Monkhorst-Pack $\Gamma$-centered
k-points mesh, which was increased to ($18\times18\times1$) for
magnetization anisotropy energies (MAEs) calculations. Partial
occupancies of wavefunctions were set according to the first-order
Methfessel-Paxton method with a smearing of 0.1 eV. As for the
considered slabs, in all relaxations we kept fixed the central most
bulklike block and allowed all other atoms to move until the forces were
smaller than 0.01~eV/\AA. The surfaces energies were evaluated from additional
calculations performed in symmetrized supercells, composed of two
equivalant surfaces on both sides of the slab and a $paraelectric$
central bulk, where Ge and Te atoms remain equidistant; same symmetric
supercells were employed in the accurate calculations of total energies
and MAEs. Dipole corrections were used for the modeling of bare
GeTe(111) surfaces.

The electronic structures and spin textures shown in the form of
projected density of states PDOS\ke maps and corresponding maps of spin
polarization $\vec{s}$\ke\ were calculated employing the GREEN
package\cite{green} interfaced with the {\it ab initio}\, SIESTA
code.\cite{siesta} For these reasons, our most stable configurations
were recalculated self-consistently with SIESTA using similar
calculation parameter values (XC functional, $k$-samplings, etc.). The
atomic orbital (AO) basis set consisted of Double-Zeta Polarized (DZP)
numerical orbitals strictly localized by setting the confinement energy
to 100~meV. Real space three-center integrals were computed over
3D-grids with a resolution equivalent to 1000~Rydbergs mesh cut-off. The
fully-relativistic pseudopotential (FR-PP) formalism was included
self-consistently to account for the SOC.\cite{soc} The electronic and
spin structures for the semi-infinite surfaces have been computed
following Green's functions matching techniques following the procedure
described in Refs \onlinecite{ysi2,loit, 2dmaterials, banis2}.

To experimentally support the calculations, the chemical interaction between Te and Fe has been monitored by X-ray Photoemission Spectroscopy (XPS), as reported in the Supplementary Information.\footnote{See Supplemental Material at URL for the experimental results.} Photoelectrons were excited using an Al-K$\alpha$ x-ray source ($h\nu$= 1486.67 eV) and analyzed through a 150 mm hemispherical energy analyzer Phoibos 150 (SPECS$^{\textsc{TM}}$), yielding an acceptance angle of ~6$^{\circ}$, a field of view of ~1.4 mm$^2$.

\begin{figure}[ht!]
\includegraphics[width=\columnwidth]{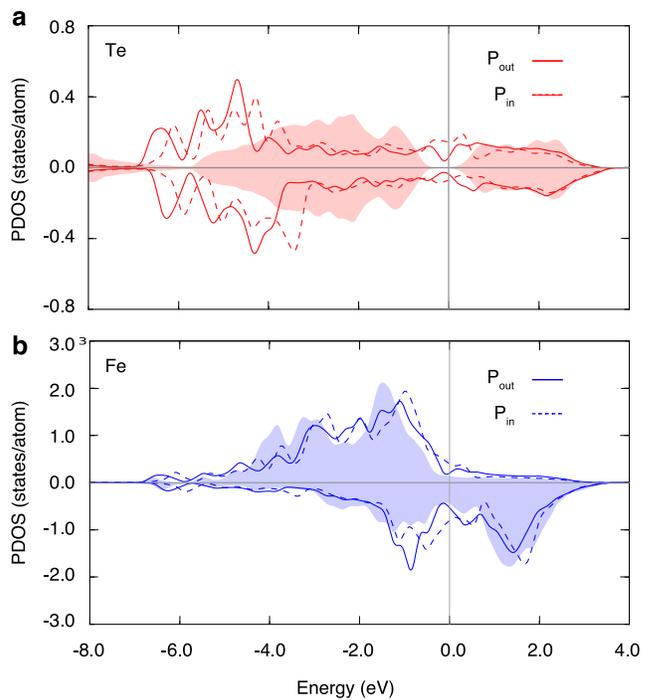}
\caption{\label{pdos}
(a) Density of states projected on interfacial (a) Te and (b) Fe atoms calculated in 6ML-Fe/GeTe(111) slab without including spin-orbit coupling. Spin majority (minority) is shown in upper (lower) panel. The solid (dashed) lines correspond to \pout\ (\pin) surface, while the shaded area denotes the PDOS of the bulklike atoms; we report in (a) the Te atom in the middle of the slab (bulk $\alpha$-GeTe phase), in (b) the atom in the middle of Fe multilayer.}
\end{figure}

\section{Results and discussion}
\subsection{Structural, electronic and magnetic properties}

\begin{figure*}[ht!]
\includegraphics[width=\textwidth]{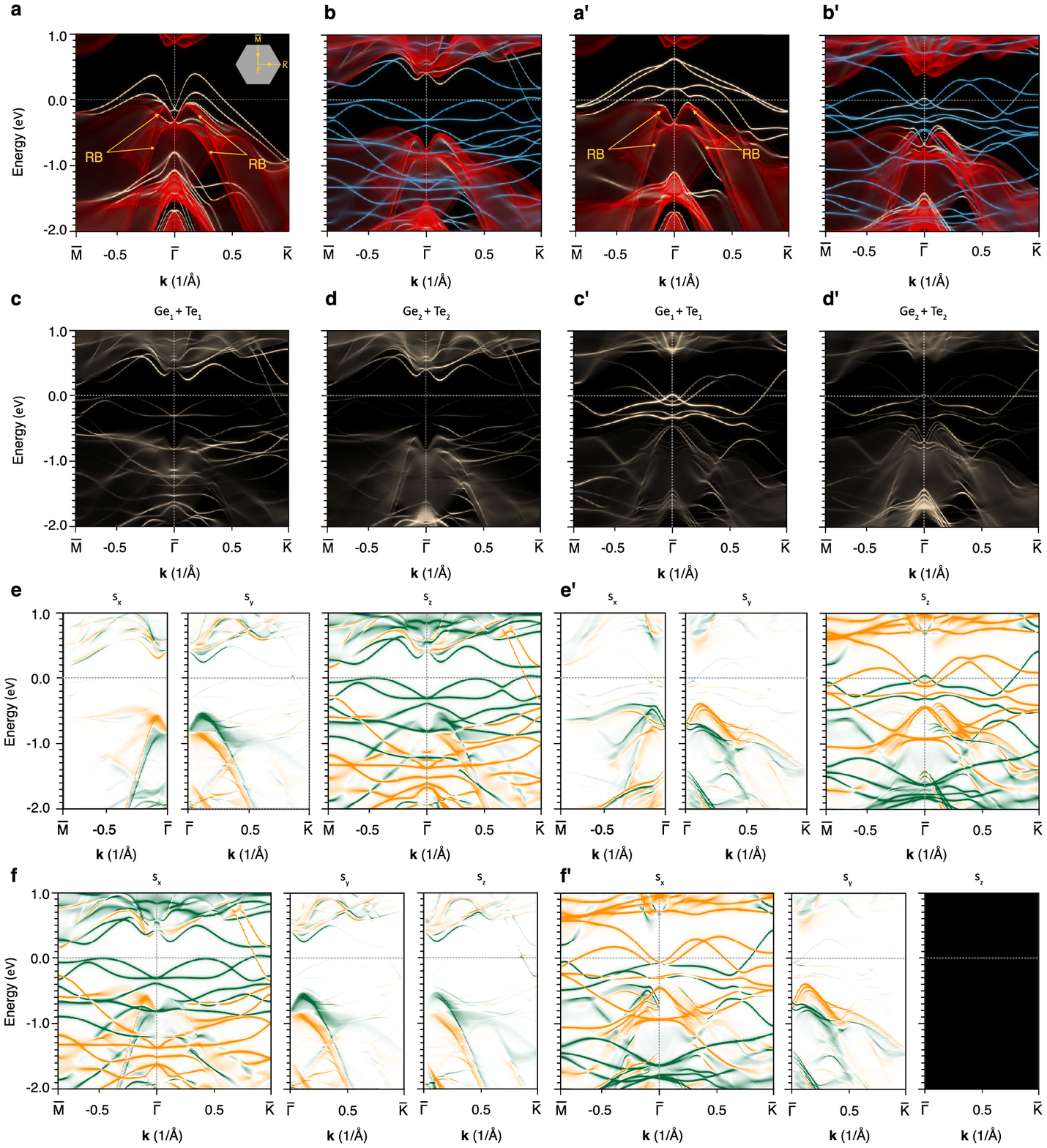}
\caption{\label{textures}
(a) Momentum and energy-resolved density of states projected on surface and bulk principal layers of the bare GeTe(111) \pout\ surface calculated within semi-infinite model via Green's functions method. The red shades represent the bulk continuum of states while the white lines correspond to purely surface bands. The yellow arrows indicate the folded bulk Rashba bands. The inset shows the Brillouin zone and high-symmetry points of hexagonal surface unit cell. (b) Same as (a) for 3ML-Fe/GeTe(111). The main color scheme same as in (a); the projections on iron atoms are additionally highlighted in blue. (c) Density of states analogous to (b), but projected only at first topmost Te and Ge atoms at the surface. (d) Same as (c), but for second layers of Te and Ge atoms. (e) Spin texture corresponding to the density of states displayed in (b) assuming QA perpendicular to the surface. The left-hand, middle and right-hand panel represent its three components, $s_{x}$, $s_{y}$, $s_{z}$, respectively. Since $s_{x}$ ($s_{y}$) achieve non-negligible values only along $\Gamma-M$ ($\Gamma-K$), the perpendicular $\Gamma-K$ ($\Gamma-M$) lines are omitted. The orange (green) shades correspond to positive (negative) values of spin polarization density. (f) Same as (e) for the QA set in-plane along $x$ axis. (a'-f') Same as (a-f) calculated for \pin\ surfaces.
}
\end{figure*}

\begin{figure*}[ht!]
\includegraphics[width=\textwidth]{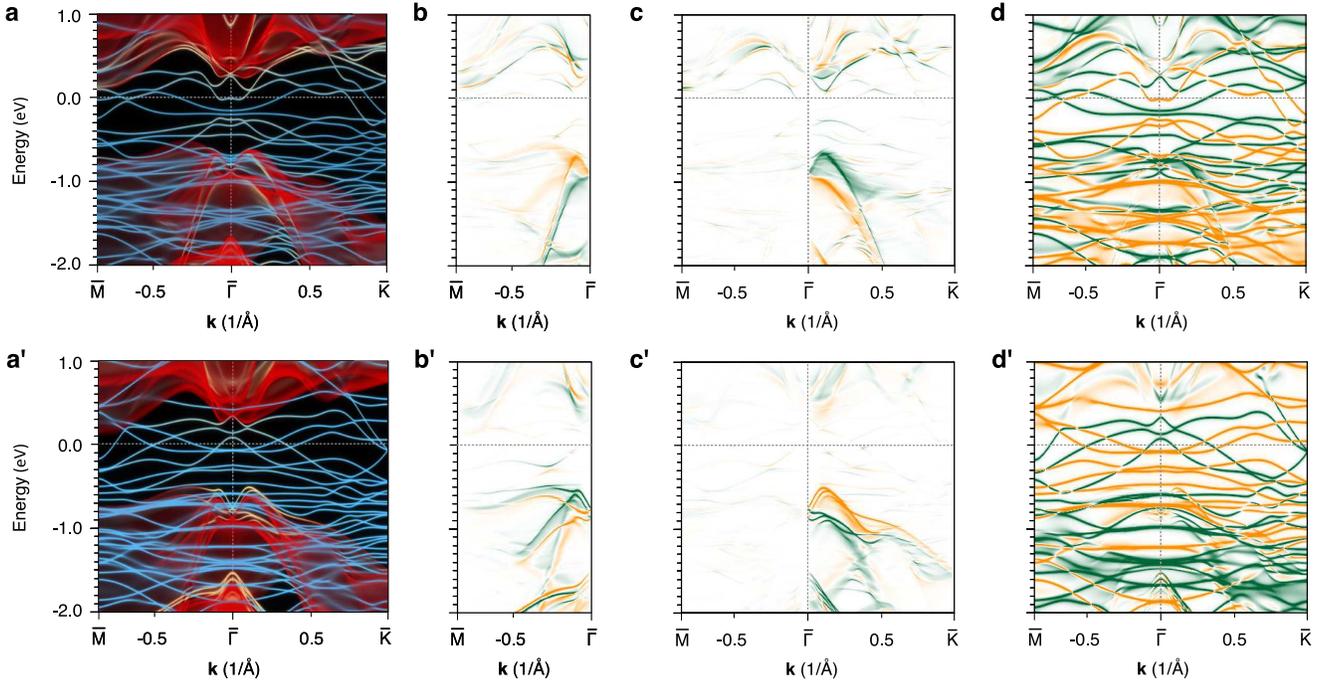}
\caption{\label{6ml}
(a) The electronic structure of 6ML-Fe/GeTe(111) calculated within semi-infinite surface model for \pout\ surface. (b-d) Spin texture projected on $x$,$y$ and $z$ axis, respectively. In (b) the $\Gamma-K$ direction is omitted, because the spin texture was found to be zero. The QA was set perpendicular to the surface. Color scheme same as in Fig.\ref{textures}. (a'-d') Same as (a-d), but calculated for \pin\ surface.
}
\end{figure*}

\setlength{\tabcolsep}{0.9em} 
\begin{table}
  \centering
  \def\sym#1{\ifmmode^{#1}\else\(^{#1}\)\fi}%
  \begin{tabular}{l*{4}{c}}
    \toprule
    & \multicolumn{2}{c}{\pout}  & \multicolumn{2}{c}{\pin} \\
        \cmidrule(lr){2-3}\cmidrule(lr){4-5}
    & \multicolumn{2}{c}{MAE = -0.43 meV}  & \multicolumn{2}{c}{MAE = -0.73 meV} \\
    \cmidrule(lr){2-3}\cmidrule(lr){4-5}
    atom\;   &  M$_S$    &  L[001]     &  M$_S$      &  L[001]       \\
    \midrule
    Ge$_1$   &  -0.01    &  0.00     & 0.00     &   0.00   \\
    Te$_1$   &  -0.02    &  0.00     &-0.03     &   0.00   \\
    Fe$_1$   &   2.06    &  0.06     & 2.18     &   0.08   \\
    Fe$_2$   &   2.62    &  0.06     & 2.60     &   0.06   \\
    Fe$_3$   &   2.32    &  0.06     & 2.40     &   0.07   \\
    Fe$_4$   &   2.70    &  0.06     & 2.68     &   0.06   \\
    Fe$_5$   &   2.56    &  0.07     & 2.63     &   0.07   \\
    Fe$_6$   &   2.82    &  0.08     & 2.82     &   0.08   \\
%
%
    \bottomrule
  \end{tabular}
\caption{Magnetic anisotropy energy (MAE), spin magnetic moments and orbital moments of the topmost surface atoms calculated for \pout\ and \pin\ phases. MAEs are evaluated as (E[001]-E[100]) per surface unit cell, Te$_{1}$ and Ge$_{1}$ refer to the interfacial surface atoms, while Fe$_{n}$ with $n$=1-6 denote iron atoms stacked as shown in Fig.1 with Fe$_{1}$ denoting the one closest to the GeTe surface. The magnetic moments are expressed in $\mu_B$.}
\end{table}

Figures \ref{struct} (a-c) and (a'-c') show the most stable geometries
for \pout\ and \pin\ surfaces, respectively. Since bare GeTe(111)
surfaces have been already studied in our previous works, their
structures are shown here only for comparison with Fe/GeTe(111)
interfaces. We have omitted the geometries of the simplest cases of 1ML
and 2ML (both are included in the Supplementary Material), because
they are clearly unlikely to be used in real devices, where the metallic
contacts for spin injection require stable ferromagnetic films of
several layers which ensure preservation of the magnetic moments. We
briefly note that the case of 1ML-Fe reveals a strong preference of the
atoms to interdiffuse into the subsurface; in fact we found such
behaviour for two most stable among three studied stacking
configurations, and for both \pout\ and \pin\ surfaces. Such tendency
can be attributed to the fact that the lattice constant of GeTe is large
enough to allow Fe atoms to fit easily below the surface, especially
when adsorbed at the $fcc$ or $hcp$  sites of GeTe(111) surface.
Certainly, the geometry of GeTe containing buried Fe atoms induces a
strong reorganization of the electric dipoles close to the surface,
leading to changes of the electronic structures including a
partial suppression of the bulk Rashba bands. Our calculations revealed
a similar interdiffusion also for two out of six studied 2ML-Fe/GeTe configurations (see SM). Similarly to the case of 1ML-Fe/GeTe, the initial configurations with Fe atoms at $hcp$ and $fcc$ sites
clearly preferred to interdiffuse, while those containing Fe atoms in $top$ configurations seem to be protected from
such structural reorganization, most likely because it would require also an in-plane shift of the adatom. This tendency explains
the lack of interdiffusion in the 3ML-Fe/GeTe slabs, as in bcc stacking in our high-symmetry models
at least one of the three Fe atoms must occupy the $top$ site. Remarkably, we have found very similar trend of interdiffusion
in analogous 1ML-Co/GeTe and 2ML-Co/GeTe indicating that the final GeTe(111) reconstruction critically depends on the exact positions of the adatoms.

As a matter of fact, the XPS investigation of chemical properties at the
Fe/GeTe interface indicates a clear tendency to interdiffusion. This is
seen already in thin films of Fe grown on GeTe at room temperature (RT) by
molecular beam epitaxy, and the phenomenon is enhanced by annealing at
200$^\circ$ C. Even though the experiments have been performed on 3 nm
thick Fe layers, as at the ultralow coverages considered in this paper
an island growth has been observed, the XPS results qualitatively
confirm the theoretical trend. Of course, real films studied at RT are far more
complex than the ideal systems used for the simulations, with
defects and vacancies largely affecting the interdiffusion. Simulations of such systems would require significantly larger supercells; such detailed structural analysis is beyond the scope of this paper.

As it can be noticed from Fig. \ref{struct} (b-b'), in 3ML-Fe/GeTe the iron atoms
are not found any more to diffuse in the subsurface, although the
geometries of the interface still reveal some peculiarities which emerge
due to the ultra-thin character of the capping layers. For example,
while the relaxations of the \pout\ side of the slab performed for
different stacking orders of Fe lead to several metastable final
geometries, the \pin\ surfaces always ends up in the configuration
presented in panel (b'), mainly because the 3ML stacking order is
removed in this case. Such behaviour can be clearly excluded in case of
thicker films, as it will be shown below.

Noteworthy, the presence of Fe not only allows for the existence of stable \pin\ termination, but even
makes this configuration more favorable ($+1.19\times 10^{-2}$
eV/\AA$^{2}$) compared to the \pout\ surface. We attribute its stability
to a formation of a strong bond between Fe and topmost Te layer which
compensates an unfavorable breaking of the short bond at the \pin\
surface. Finally, Fig. \ref{struct} (c-c') show the structural
properties of the most stable 6ML-Fe/GeTe(111) configurations. Although
the GeTe surfaces remain roughly the same as in case of capping with
3ML-Fe, the ferromagnetic layers adapt different geometries;  the
preferred stacking order is different than in the 3ML-Fe/GeTe(111), and
identical for the \pout\ and \pin\ models. Interestingly, both \pout\
and \pin\ surfaces reveal shorter adsorption distances, which is better
captured in panel (d) where all the interlayer distances of
considered interfaces are summarized. Finally, we note that in
6ML-Fe/GeTe(111) all initial Fe configurations for both polarization
phases preserved their stacking after relaxation; this can be
intuitively explained by the fact that the structure of 6ML-Fe already
approaches a crystalline one, thus preventing any severe re-ordering of
the outer layers. Again, the \pin\ configuration was found to be
significantly more stable than \pout\ ($+1.14\times 10^{-2}$
eV/\AA$^{2}$).


Further insights on Fe/GeTe(111) interfaces have been gained by performing the calculations of MAEs due to interfacial magnetocrystalline (single-ion) anisotropy, neglecting dipolar contributions; the corresponding values are listed in Table I for 6ML Fe coverage. For both directions of $\vec{P}$, magnetocrystalline anisotropy favors a perpendicular-to-plane configuration of the Fe magnetic moment (MM). On the other hand, the \pin\ configuration reveals a notably larger magnetocrystalline MAE (by as much as 0.3 meV) which confirms the existence of a magnetoelectric coupling in the interfaces with thin Fe layers.
The MAE dependence on the Fe thickness is a delicate issue. It is  reported in the literature that the single ion anisotropy in pure iron thin films strongly oscillates with the number of Fe layers up to quite large thicknesses.\cite{li, shape} In our Fe/GeTe case, for Fe thicknesses larger than 6 ML, the simulations become too expensive from the computational point of view and  results  with the required accuracy cannot be reported.  However, the simulations of 8ML-Fe/GeTe(111) and 10ML-Fe/GeTe(111) confirmed that magnetocrystalline anisotropy favors the perpendicular-to-plane configuration.

The impact of interfacial magnetocrystalline MAE on the real arrangement of Fe magnetization can be understood by comparing it with the magnetostatic energy term responsible for shape anisotropy. As previously reported by Bornemann \textit{et al.},\cite{mae_minar} for small Fe thickness the dipolar energy can be estimated by using the classical concept of magnetostatic energy,\cite{shape} which quantitatively reproduces the quantum mechanical results. For a thin film, the volume magnetostatic energy density can be written as
\begin{equation}
E_{M}= 1/2\mu_{0}M_{S}^2\cos^{2}\theta
\end{equation}
where $\mu_0$ is the vacuum permittivity, $M_{S}$ is the saturation magnetization and $\theta$ is the angle between the sample magnetization and the out-of-plane direction.  The shape anisotropy density per surface unit cell, to be compared with the MAE values reported in Table I, can be calculated multiplying $E_{M}$ by the volume of the unit cell. This is given by the product of the area of the surface unit cell ($A=15.45$ \AA$^{2}$ as the hexagonal cell of GeTe has a lattice parameter of 4.22 \AA\, and 3 Fe atoms per cell) by the average layer spacing in bcc-like Fe/GeTe along the pseudocubic [111] direction (about 0.7 \AA) multiplied by the number of layers ($n$). For the case of 6ML considered in Table I, assuming a Fe bulk saturation magnetization $M_{S}= 1.74\cdot10^{6}$ A/m, we obtain a shape anisotropy energy density per unit cell of 0.77 meV. This value is very close to that of single-ion MAE for the \pin\ polarization and larger than that for \pout, thus indicating that for 6ML the large change in MAE induced by polarization reversal can influence the overall anisotropy displayed by the Fe film. Ultrathin Fe films can have an out-of-plane easy axis, while at larger Fe thickness the volume magnetostatic contribution largely exceeds the single-ion MAE, which is confined at the interface, and the magnetization reorients in the film plane. From the estimation above, the spin reorientation transition should take place at a critical thickness of the order of 6 ML, corresponding to about 0.42 nm. This is fully consistent with our previous result showing that 5 nm of Fe on GeTe(111) display a clear in-plane hysteresis loop.\cite{fert}

In addition, Table I reports the values of MMs calculated for the surface
atoms, including the orbital moments obtained for the [100] magnetic
orientation. Any differences between \pout\ and \pin\ configurations can
be noted mainly at the Fe atoms located close to the semiconductor; the
interaction between Fe and Te seems to be responsible for the
appreciable reduction of Fe MMs which experience a sizeable decrease (of
the order of 0.1 $\mu_{B}$) when changing from \pin\ to \pout\ surface.
We emphasize that in both cases the interfacial Te atom reveals a small
magnetic moment (0.02-0.03 $\mu_B$) antiferromagnetically coupled to
that of Fe. When inspecting the DOS projected on interfacial Te and Fe
presented in Fig. \ref{pdos} we can indeed observe a strong
hybridization of Fe and Te states within the whole considered energy
window, including the region close to the Fermi energy, where the Fe
$3d$ states induce both spin majority and minority in the DOS of Te. X-ray photoemission data reported in the Supplementary Information support the existence of a
preferential interaction between Fe and Te. While Ge peaks do not move
in energy upon formation of the Fe/GeTe interface, Te display a
core-level shift towards higher binding energy, compatible with that
reported in case of Fe films deposited on Bi$_2$Te$_3$.\cite{ivana} In
closer details, while the presence of Fe induces new states, the GeTe gap
decreases and makes both interfaces conducting; for \pout\ there is a sort
of \textit{pseudo-gap} very close to the Fermi energy, whereas for \pin\
the metallic behaviour becomes robust. The value of DOS projected on the
interface Te atom increases by $\sim$ 2.5 times at $E_{F}$ when changing
from \pout\ to \pin, a result which might have important consequences
for any spin-injection related process or exploitation for ME junctions.


\subsection{Electronic structures and spin texture}
Figures \ref{textures} (a-a') show the GeTe(111) band structures
calculated in the form of projected density of states PDOS\ke for each
polarization configuration; the surface and bulk projections are
distinguished by using white and red shades, respectively. The folded
bulk Rashba bands are indicated by the arrows.  Next, we present in panels (b-b'),
side-by-side, the analogous electronic structure maps
calculated for the 3ML-Fe/GeTe(111). Although the geometry of the
surface is hardly affected compared with bare GeTe (see the interlayer
distances in Fig.\ref{struct}d), the influence of Fe on the band
structure is indeed huge. In particular, the surface states (SS) are
completely removed at the \pout\ side, and strongly suppressed at the
\pin\ surface due to the several Fe states residing inside the bulk gap
(highlighted in blue in the maps). Similar intense states cover
practically the whole displayed energy region, but without affecting the
most relevant bulk Rashba bands. In order to gain further insights on
the screening properties of GeTe with respect to interface electronic
states, we additionally present, in panels (c-d) (correspondingly c'-d'
for the \pin\ surface), the density of states projected only on topmost
surface atoms, {\it i.e.} first and second Te-Ge bilayers. Certainly, the
projection on the two topmost atomic layers (Ge$_1$+Te$_1$) reveals the presence of Fe-induced
states, which points for a strong hybridization at the
interface. However, these bands fade out quite rapidly with the depth;
at the third and fourth atomic layers (Ge$_2$+Te$_2$) we can observe
only weak traces of few of them. Instead, the bulk Rashba
bands are already clearly visible, showing that interface states are
efficiently screened by GeTe, consistent with its semiconducting
behaviour. It is worthwhile to note that different results were found by Krempask\'{y} \textit{et al.} in an apparently similar multiferroic system Ge$_{(1-x)}$Mn$_x$Te where the structure of $bulk$ bands depends on Mn concentration.\cite{krempasky, krempasky_prx} In particular, it was found that the bulk Rashba bands possess a Zeeman gap between the Dirac points, whose presence is attributed to rather strong exchange interaction and its interplay with SOC. Our results do not reveal such effect in Fe/GeTe due to the fact that Fe induces changes mainly at the surface of GeTe(111), in contrast to Ge$_{(1-x)}$Mn$_x$Te where magnetic impurities are homogeneously distributed in the sample. In fact, even in case of stronger interaction (such as interdiffusion in 1ML-Fe/GeTe), we have not found any traces of Zeeman gap.

Panels (e-e') display the corresponding spin-resolved density of states,
$\vec{s}$\ke calculated for the quantization axis (QA) normal to the
surface, which was found to be the most stable one (see Table I). We
visualized the spin textures separately for three components $s_x$,
$s_y$ and $s_z$. In the case of in-plane projections, we omitted the
directions of the BZ along which the spin texture was negligible. These
directions are consistent with the expected Rashba-like spin-momentum
locking, i.e. the spin-components are found to be non-zero only when
perpendicular to the momentum. Expectedly, the strongly spin polarized
iron bands manifests mainly in the $s_z$ component parallel to the QA,
overlaying the still visible bulk states, while the in-plane projections
$s_x$ and $s_y$ reveal mainly the spin texture of bulk bands, hardly
modified by the interaction with Fe. Setting the QA along $x$ ($y$) (f-f' ) yields a similar scenario, with spin textures of Fe clearly dominating the $s_x$ ($s_y$) components, and purely bulk Rashba bands
manifesting in the complementary projections of the $\vec{s}$. This
shows that the hybridization does not strongly depend on the QA.
Finally, the electronic/spin properties of 6ML-Fe/GeTe(111) reported in
Fig.\ref{6ml} resemble those calculated for 3ML-Fe/GeTe(111); the only
differences are several new Fe states well visible in PDOS, but hardly
interacting with the bulk continuum.
This confirms the robustness of the
interface electronic structure, both with respect to the Fe thickness
and to the stacking order.

Overall, our electronic and spin structure calculations show that the
Fe/GeTe(111) interface, in general, produces strong interface
hybridization but leaves the bulk Rashba bands hardly altered already at
the sub-surface level, which seems promising for their further
exploration and exploitation. Recent spin-pumping experiments have
indeed revealed the SCC in this system, which could originate from
interface or bulk Rashba states, according to the inverse Edelstein or
Inverse Spin Hall effects, respectively. \cite{iee_fert, iee, iee_apl}
Our results shed light on this subject, as we have seen that the
creation of the Fe/GeTe interface tends to suppress surface Rashba
states. Thus, we suggest that SCC phenomena in
this system could be mainly related to bulk Rashba states, whose
dispersion and spin character is almost unaffected by the presence of
the Fe/GeTe interface. On the other hand, DFT calculations indicate that
the creation of the Fe/GeTe interface has a deep impact on the GeTe
bandstructure at the interface. Starting from the typical band line-up
of a $p$-doped material, consistent with the large concentration of Ge
vacancies in real films, the bulk Rashba states in the valence band
shift downwards by about 0.5 eV and the Fermi level moves towards the
centre of the gap. In these conditions, spin transport at the interface
is expected to involve also states from the conduction band, having
different Rashba parameters and thus possibly leading to a different
behavior with respect to that expected in case of $p$-doped GeTe. A
detailed explanation of the mechanism, including determining the exact
role of bulk or/and interface states would require additional out-of
equilibrium spin-transport calculations, which are, however, beyond the
scope of the present paper. On the other hand, in analogy with previous
works,\cite{fert} our results point to the crucial
role of the interface between a ferromagnet and a Rashba material in
determining the spin transport properties. The engineering of the
interface, by properly choosing the ferromagnet and/or by inserting an
intermediate layer, provides an additional degree of freedom to optimize
spin-dependent effects like SCC. This calls for
further theoretical and experimental investigations of this system.

\section{Summary}
In summary, we have performed a detailed analysis of multilayer Fe films
deposited on $\alpha$-GeTe(111) surfaces. First, we have revealed that
the Fe capping layers stabilize the GeTe surfaces with the two different
polar configurations close to the surface, with the electric dipole
pointing outwards and inwards, in contrast to bare GeTe surfaces, where
the latter is unstable. The ultrathin Fe thicknesses (1ML and 2ML)
modify the structure of GeTe(111), consistently with the
experimental results pointing to a large interdiffusion of Fe ions
within the GeTe substrate. However, starting from 3ML the topmost surface
atoms in GeTe remain hardly affected, indicating that for any practical
purposes rather thick ferromagnetic films should be employed. Finally, we unveiled
the electronic structures and spin textures,
including the effects of both directions of $\vec{P}$ and different thicknesses of the
Fe overlayer. In all cases the Fe states strongly hybridize with the GeTe surface,
leading to a suppression of the Rashba surface states. Importantly, the
bulk Rashba bands remain almost electronically unaffected and are only
altered at the interfacial GeTe layer, consistently with the expected good
screening properties of GeTe.

In conclusion, our theoretical and experimental work paves the way for the understanding
of the microscopic mechanisms at the heart of potentially useful new generations of interfaces.
The key idea of combining ferromagnetic overlayers with active ferroelectric
Rashba semiconductors may grasp the avenue to engineer groundbreaking spintronics
devices by making use, for example, of the already proven efficiency of Fe/GeTe heterostructures.\cite{fert}

\begin{acknowledgments} We are grateful to Dr. J. I. Cerd\'{a} for helpful comments on the calculation strategy in SIESTA. The work at CNR-SPIN was performed within the
framework of the Nanoscience Foundries and Fine Analysis (NFFA-MIUR Italy) project. This work has been supported by Fondazione Cariplo and Regione Lombardia, grant no 2017-1622 (project ECOS). The experimental work reported in the Supplementary Information has been partially carried out at Polifab, the micro and nanofabrication facility of Politecnico di Milano. D.D.S. acknowledges the German DFG through
the SFB1170 "Tocotronics" and the ERC-StG-336012-Thomale-TOPOLECTRICS. Part of the calculations has been performed in CINECA Supercomputing Center in Bologna.
\end{acknowledgments}

%
\end{document}